\documentclass[a4paper]{ISOStyle}
\usepackage{epsfig}

\begin{document}
\voffset-.3in

\title{Galaxies:  The Long Wavelength View}

\author{J. Fischer\inst{1}} \institute{Naval Research Laboratory,
  Remote Sensing Division, Washington, DC, USA }

\maketitle 

\begin{abstract}

Far-infrared spectroscopy obtained with the ISO LWS has shown that there is 
strong variation (more than 2 orders of magnitude) in the [\ion{C}{II}]/FIR ratios in galaxies extending from blue compact dwarfs, to normal and starburst galaxies, down to elliptical and ultraluminous galaxies (ULIGs).  The variation in the relative [\ion{C}{II}] 158 $\mu$m line strength has been attributed to low metallicity in blue compact dwarfs, high $\langle$$G_o$$\rangle$/$n$ for normal galaxies and ULIGs, soft radiation fields in ellipticals, and extinction or enhanced abundance of dust in ionized regions in ULIGs. 

Full ISO/LWS far-infrared (43 - 197 $\mu$m) spectra of six nearby IR-bright galaxies reveal a dramatic progression of decreasing fine-structure line emission from ionized species to molecular (OH and H$_2$O) absorption line dominated.  The archetypical ULIG, Arp~220, lies at the absorption line dominated end of this sequence.  For Arp~220, radiative transfer models indicate that it is optically thick in the FIR and that the water molecules observed in absorption are radiatively excited.  If extinction plays a role in the sequence it appears from this analysis that the affected regions are heavily obscured even in the far-infrared, while the \emph{detected} line emission is not more obscured in ULIGs than in starbursts.  Linear correlation between polycyclic aromatic hydrocarbon (PAH) 6.2 $\mu$m feature strength and the [\ion{C}{II}] 158 $\mu$m line strength in starbursts and ULIGs suggests a similar effect for these emitting species, and that the detected PAH emission is not more obscured in ULIGs than in lower luminosity starbursts. 

\end{abstract}

\section{INTRODUCTION}

Far-infrared spectroscopy obtained with ISO's Long Wavelength
Spectrometer (LWS) probes the global conditions in the ionized,
molecular, and dust components of the interstellar medium in galaxies.
Indeed, the great diversity in the FIR spectra of galaxies is
indicative of varying gas and dust content, location, and excitation
along the line of sight.  Far-infrared spectroscopy is not a sensitive
probe of the hard radiation fields characteristic of AGN.  Instead, the far-infrared spectra of galaxies are often dominated by emission from the strongest cooling lines (i.e. [\ion{C}{II}] 158 $\mu$m and [\ion{O}{I}]63,146 $\mu$m) of the photodissociation regions (PDRs) from which the far-infrared continuum
emission is thought to arise and by the forbidden fine-structure lines
from H~II regions.  Because it is much less affected by extinction due
to dust than spectroscopy at shorter wavelengths, FIR spectroscopy can
be an important tool in probing the starburst properties of the highly
obscured regions often inferred to be present in infrared-bright galaxies.

In the first part of this review, I will describe the results of
far-infrared spectroscopic surveys of a variety of morphological types
of galaxies, including dwarf, normal, starburst, and ultraluminous
galaxies (ULIGs).  The far-infrared spectra of active galactic nuclei are 
discussed by Spinoglio et~al. elsewhere in these proceedings.  I will 
then discuss the still uncertain physical conditions and evolutionary 
effects responsible for the weak emission line galaxies, many of which 
are ULIGs.  Where possible I will discuss cross-instrument studies and 
compare the results of mid- and far-infrared spectroscopic surveys.

\section{H~II REGION AND PDR DIAGNOSTICS}
\label{JF_sec:hiipdrs}

The LWS range includes seven diagnostic fine-structure lines:  [\ion{C}{II}] 158 $\mu$m, [\ion{O}{I}] 63 and 146 $\mu$m, [\ion{N}{II}] 122 $\mu$m, [\ion{O}{III}] 52 and 88 $\mu$m, and [\ion{N}{III}] 57 $\mu$m.  The [\ion{N}{III}] 57 $\mu$m /[\ion{N}{II}] 122 $\mu$m line intensity ratio is a sensitive indicator of the hardness of the UV radiation field when T(eff) $\sim$ 33,000 K.  This pair thus indicates the upper mass limit (or, age) of any purported starburst.  The [\ion{O}{III}] 52,88 $\mu$m lines can be used to derive the density for moderate densities of the ionized gas and together with the [\ion{N}{III}] 57 $\mu$m and [\ion{N}{II}] 122 $\mu$m lines can be used to estimate the average O/N abundance ratio, and hence, star formation history in the galaxy.  The [\ion{N}{II}] 122 $\mu$m line can also be used to discern the fraction of [\ion{C}{II}] emission which arises from low density ionized gas.  

\begin{figure}[tl]
  \begin{center}
    \epsfig{file=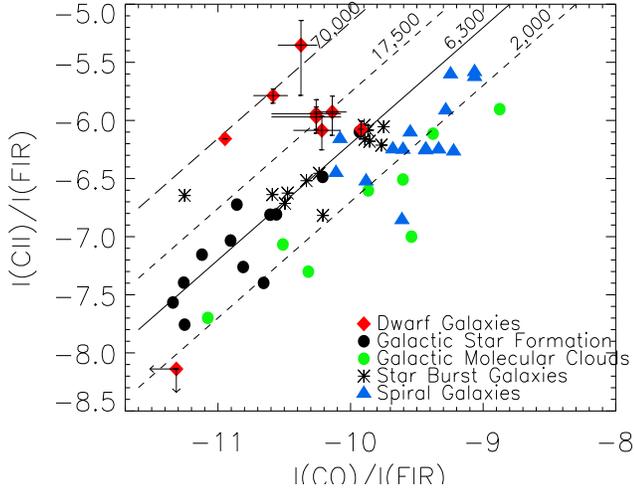, width=8.5cm}
  \end{center}
\caption{A comparison of low-metallicity galaxies with galactic star-forming 
regions and normal and starburst galaxies. Lines of constant I[C II]/I(CO) ratios are shown and range from ~2000 up to 70,000 for some dwarf galaxies.  The ratios of both axes are normalized to the local 
interstellar radiation field (1.3 $\times$ 10$^{-4}$ erg s$^{-1}$ cm$^{-2}$ 
sr$^{-1}$).  From Madden (2000). \label{JFfig_dwarfs}}
\end{figure}

\begin{figure}[!lbh]
  \begin{center}
    \epsfig{file=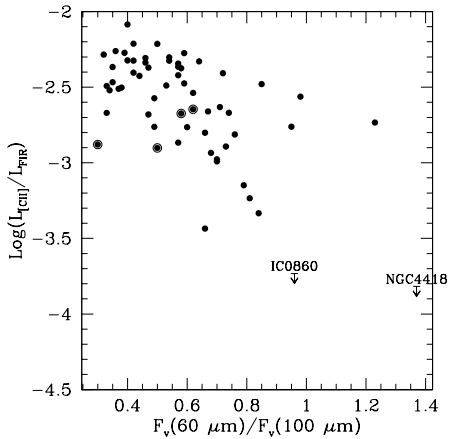, width=6.0cm}
  \end{center}
\caption{The [C II]158/FIR flux ratio versus $F_{60}/F_{100}$ from IRAS.  Early-
type galaxies are denoted by bullseye symbols.  From Malhotra et~al. (2000a). \label{JFfig_ciifira}}
\end{figure}

By definition, only photons less energetic than those capable of ionizing 
hydrogen (i.e. with energies less than 13.6 eV) pass the ionization boundary of 
an HII region and heat the gas within PDRs.  The most energetic of these, the far ultraviolet photons with energies of $\sim$ 6 - 13.6 eV are absorbed by dust followed by the ejection of  energetic photoelectrons which then heat the gas in the ensuing layers of H~I, [\ion{C}{II}] (ionization energy 11.3 eV), [\ion{O}{I}] and molecules.  Models of the heating and cooling of the gas and dust, including polycyclic aromatic hydrocarbon (PAH) heating, have produced diagnostic diagrams relating the strong cooling line emission and the total far-infrared dust emission to the average radiation density and hydrogen density in the PDR (\cite{JF_wolf90}; \cite{JF_kauf99}).   These models can be applied directly if both the total FIR and the line emission come from the PDR and if the line emission is not affected by absorption due to foreground gas and dust.  Since the [\ion{C}{II}]158 $\mu$m and [\ion{O}{I}]63,145 $\mu$m, lines from these regions are very strong, they can be used to trace the conditions in dusty star forming regions, even in elliptical and blue compact dwarfs where the amount of dust is relatively low. 

\section{THE EFFECTS OF METALLICITY}
\label{JF_sec:lowmet}

The metallicity of a galaxy provides information about its star 
formation history.  Moreover, the measurement of the metallicities of 
galaxies at high-z will provide clues to the star formation history of the 
early universe.  Thus the development and understanding of infrared 
metallicity diagnostics are important for future missions such as SIRTF, 
NGST, and FIRST. 

\cite*{JF_mad00} has analyzed ISO and KAO (Kuiper Airborne Observatory) 
observations of the [\ion{C}{II}] 158\,$\mu$m line in a sample of 15 dwarf galaxies with metallicities ranging from 0.1 to 0.5 solar.  She finds very
high values of [\ion{C}{II}]/CO(1-0) and somewhat higher [\ion{C}{II}]/FIR (Figure~\ref{JFfig_dwarfs}) in these low metallicity systems relative to other galaxies discussed in Stacey et al. (1991), but does not find an unambiguous direct correlation of the [\ion{C}{II}]/CO ratio with metallicity within her sample.  She interprets the enhanced [\ion{C}{II}]/CO(1-0) to be the result of reduced dust abundance, which allows the UV radiation to penetrate further into the molecular core (\cite{JF_malblack88}; \cite{JF_israe96}), producing a geometric dilution effect. 

\section{NORMAL GALAXIES}
\label{JF_sec:norm}

\begin{figure}[!br]
  \begin{center}
    \epsfig{file=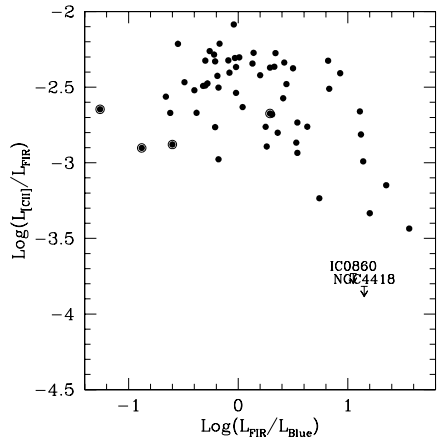, width=6.0cm}
  \end{center}
\caption{ The [C II]158/FIR flux ratio versus $F_{FIR}/F_B$.  Early-type 
galaxies are denoted by bullseye symbols.  From Malhotra et~al. (2000a). \label{JFfig_ciifirb}}
\end{figure}

\begin{figure}[!tl]
  \begin{center}
    \epsfig{file=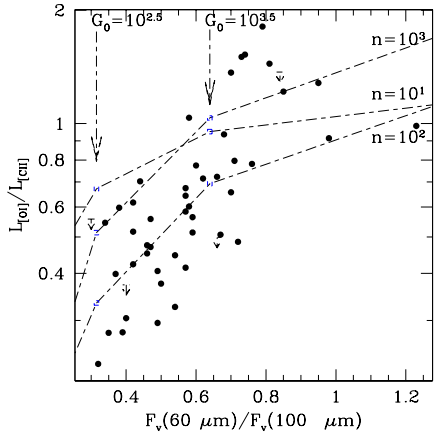, width=6.0cm}
  \end{center}
\caption{ The observed correlation between the ratio of the two PDR cooling 
lines [O I]63/[C II]158 and $F_{60}/F_{100}$ compared with the theoretical PDR models (Kaufman et~al. 1999).  Although the models are in qualitative agreement with the observed ratios, the observed ratios are lower than model predictions for about half of the sample.  
From Malhotra et~al. (2000b). \label{JFfig_oiciia}}
\end{figure}

Malhotra et al. (1997; 2000a,b) have analyzed their survey of
a diverse sample of normal galaxies in the [\ion{C}{II}]158, [\ion{O}{I}]63,145, [\ion{N}{II}]122, and [\ion{O}{III}]88,52 $\mu$m far-infrared fine-structure lines.  Since these galaxies are relatively faint in the far-infrared continuum and fine-structure lines, their work concentrates on the analysis of the strong [\ion{C}{II}]158 and [\ion{O}{I}]63 $\mu$m PDR cooling lines.  Using $F_{60}/F_{100}$ and $F_{FIR}/F_B$ as measures of the dust heating radiation density and star formation activity respectively, they argue that the inverse correlation of [\ion{C}{II}]/FIR with both of these (Figure~\ref{JFfig_ciifira} and Figure~\ref{JFfig_ciifirb}) is evidence that the low values of [\ion{C}{II}]/FIR found for warm, infrared-bright galaxies such as IC~860 ($<$ 1.96 $\times$ 10$^{-4}$) and NGC~4418 ($<$ 1.53 $\times$ 10$^{-4}$) can be attributed to lower gas heating efficiency arising from positively charged grains for high values of $\langle$$G_o$$\rangle$/$n$.  They suggest that this effect is also responsible for the low [\ion{C}{II}]/FIR found in 11 out of 12 ULIGs by Luhman et al. (1998; 2000) (see Section~\ref{JF_sec:ulig}).  This is further substantiated by a positive correlation between the ratio of the two strong PDR cooling lines [\ion{O}{I}]63/[\ion{C}{II}]158 and dust temperature (Figure~\ref{JFfig_oiciia}), as is predicted by theoretical PDR models.  

\begin{figure}[!tr]
  \begin{center}
    \epsfig{file=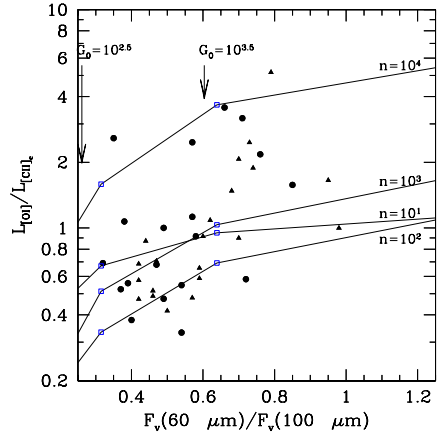, width=6.0cm}
  \end{center}
\caption{ As in Figure~\ref{JFfig_oiciia}, but here a correction is made for 
[C~II] arising from the diffuse ionized medium, based on [N II]122 observations.  The triangles denote galaxies where [N II] 2-$\sigma$ upper limits were used.  From Malhotra et~al. (2000b). \label{JFfig_oiciib}}
\end{figure}

In order to obtain quantitative agreement with the PDR models of 
Kaufman et al. (1999), they find that it is necessary to significantly correct the observed [\ion{C}{II}] fluxes for a contribution from diffuse ionized gas based on measured [\ion{N}{II}] line fluxes (Figure~\ref{JFfig_oiciib}).  With this correction they find that their data is best fit by $\langle$ $G_o$ $\rangle$ $\propto$ $n^{1.4}$ and conclude that this can be explained if the PDRs in high $\langle$$G_o$$\rangle$/$n$ galaxies are associated with embedded H~II regions and larger giant molecular clouds. 

\section{INFRARED-BRIGHT GALAXIES}
\label{JF_sec:irbgal}

Population synthesis and photoionization models of dusty, infrared-bright 
galaxies such as Arp~299, M~82, and NGC 4038/39, indicate that short-lived 
bursts of star formation with ages of 3-7 $\times$ 10$^6$ years, can best explain the LWS spectra of these galaxies (\cite{JF_saty00}, \cite{JF_colb99}, and \cite{JF_fis96}).  \cite*{JF_thor00} find similar results from an analysis of their [\ion{Ne}{III}]15.6 $\mu$m, [\ion{Ne}{II}] 12.8 $\mu$m survey  of 27 starburst galaxies.  Satyapal et al. show that the models derived from LWS measurements including all of the individual components of Arp 299 are indeed consistent with the ensemble of model fits to high spatial resolution Brackett $\gamma$ images of individual components of this galaxy. \cite*{JF_ung00} find that even in Cen A, the extended far-infrared emission is powered primarily by star formation rather than by the central obscured AGN. 

\begin{figure}[!tl]
  \begin{center}
    \epsfig{file=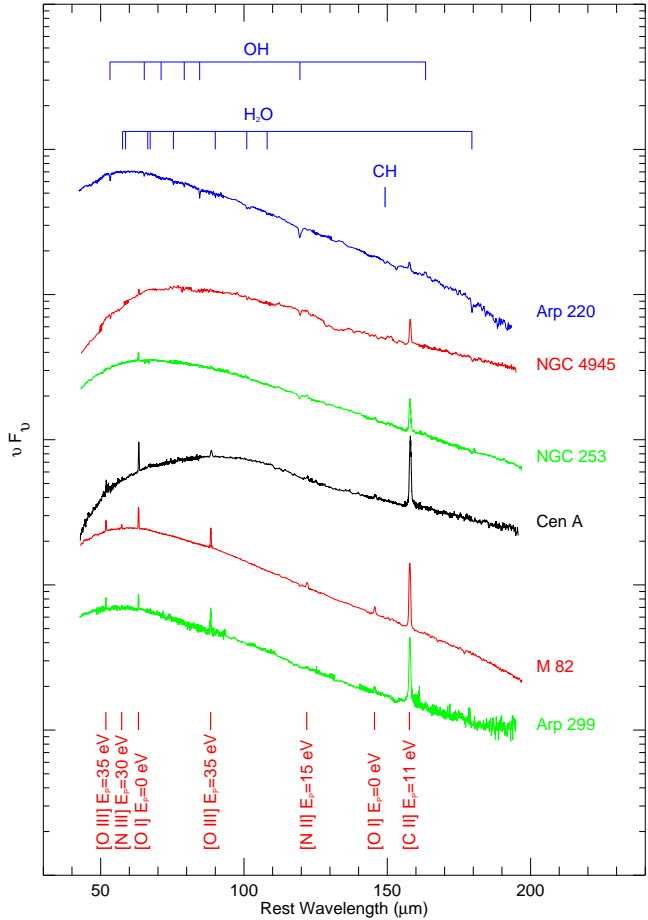, width=8.5cm}
  \end{center}
\caption{ The full ISO Long Wavelength Spectrometer spectra of six IR-bright 
galaxies.  The spectra have been shifted and ordered vertically according to the 
equivalent width of the [O III]88 $\mu$m line.  The excitation potential, the energy required to create the species, is given in eV at the bottom of the figure.  From Fischer et~al. (1999). \label{JFfig_firseq}}
\end{figure}

\cite*{JF_fis99} have presented a sequence of full, high signal-to-noise LWS 
spectra of six infrared bright galaxies (see Figure~\ref{JFfig_firseq}) in order 
of the relative strength of the [\ion{O}{III}] 52, 88\,$\mu$m fine-structure lines.  The spectral sequence extends from the strong emission line galaxies Arp 299 (Satyapal et al. 2000) and M~82 (\cite{JF_colb99}) to the ULIG Arp~220, whose spectrum is dominated by absorption lines of OH, H$_{2}$O, CH, and [\ion{O}{I}], with only very weak [\ion{C}{II}] 158\,$\mu$m and OH 163\,$\mu$m line emission.  Intermediate in the sequence are Cen~A (\cite{JF_ung00}), NGC~253 (\cite{JF_brad99}) and NGC~4945 (\cite{JF_lord00}), showing weak [\ion{O}{III}] and [\ion{N}{III}] lines while their PDR emission lines remain moderately strong.  In this sequence, the strength and richness of the molecular absorption spectra are anti-correlated with the equivalent widths of the
fine-structure emission lines.  For example, M~82 shows faint OH
absorption from the ground level at 119\,$\mu$m, while
NGC~253 shows absorption from the ground-state in three cross-ladder
transitions and an emission line cascade in the 79\,$\mu$m and 163\,$\mu$m 
lines.  In Arp~220, absorption from rotational levels
as high as 416 K and 305 K above the ground state is present for OH and
H$_2$O respectively, and the [\ion{O}{I}]63\,$\mu$m line is seen in
absorption.  

Using the temperature-insensitive [\ion{O}{III}] 52 $\mu$m/[\ion{O}{III}] 88 $\mu$m line ratio as a probe of density, Fischer et~al. find no clear
dependence of the $[\ion{O}{III}]88/F_{FIR}$ ratio on electron density and all of the measured [\ion{O}{III}] line ratios were found to fall within the range 0.6 - 1.2, consistent with electron densities between 100 - 500 cm$^{-3}$.  From this they infer that neither density nor far-infrared differential extinction
(between 52 - 88 $\mu$m) to the ionized gas appears to be the single dominant parameter in the observed sequence.  Rather, since both of the line ratios
[\ion{N}{III}]57/[\ion{N}{II}]122 and [\ion{O}{III}]52/[\ion{N}{III}]57 increase as a function of the [\ion{O}{III}]88/$F_{FIR}$ ratio, they infer that [\ion{O}{III}]88/$F_{FIR}$ correlates with excitation.  They suggest that progression to soft radiation fields or dusty H~II regions may explain these effects and that the absorbing molecules may be located in dense photodissociation regions, where they could be excited radiatively by the far-infrared emission from warm dust.  If extinction plays a role in this sequence it appears from this analysis that the affected regions are very heavily obscured even in the far-infrared, while the detected line emission is relatively unobscured.  In this case, the progression to low excitation could be a result of total obscuration of the youngest starburst population and/or the central AGN. 

\section{ULIGS AND WEAK EMISSION-LINE GALAXIES}
\label{JF_sec:ulig}

\begin{figure}[tl]
  \begin{center}
    \epsfig{file=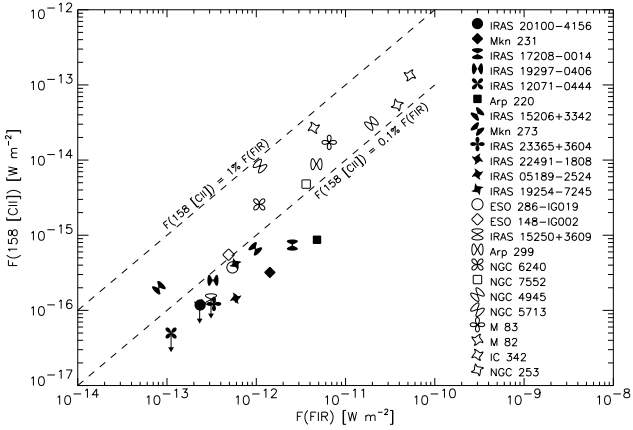, angle=90, width=8.5cm}
  \end{center}
\caption{ The [C II] 158\,$\mu$m line flux versus FIR flux for 12 ULIGs
observed with the LWS (Luhman et~al. 1998; 2000) compared with a sample of normal and starburst galaxies (Luhman et~al. 1998; Lord et~al. 1996; Colbert et~al. 1999; Stacey et~al. 1999; Bradford et~al. 1999).  In the symbol key, the galaxies are listed in order of luminosity from top to bottom and the ULIG symbols are black filled, while the lower luminosity galaxies are unfilled.
ULIGs are defined as in Sanders \& Mirabel (1996).  The dashed lines mark the regime typical of normal and starburst galaxies.  From Luhman et~al. (2000). \label{JFfig_ciifir}}
\end{figure}

The full LWS spectrum of the second brightest ultraluminous galaxy
Mkn~231 (\cite{JF_harv99}) is similar to that of Arp~220 (to
within the achieved signal-to-noise ratio).  It is dominated by
OH absorption, with similar OH absorption line ratios, and only
faint [\ion{C}{II}]158\,$\mu$m line emission is present.  Based on the mid-infrared spectra of a sample of nearby ULIGs, \cite*{JF_genz98} infer that
Mkn~231 has a strong AGN component while the far-infrared luminosity of
Arp~220 is powered by a starburst.  Thus the similarity of the
far-infrared spectra of these two ultraluminous galaxies is 
surprising.  Indeed, Luhman et~al. (1998; 2000) find that the [\ion{C}{II}]158/FIR ratio in ULIGs is typically an order of magnitude fainter than in lower luminosity starburst and normal galaxies
(Figure~\ref{JFfig_ciifir}).  

Based on single dust temperature fits and measured millimeter size constraints 
the far-infrared dust emission of the nearest ULIG, Arp 220, is thought to be
optically thick (\cite{JF_sco91}; \cite{JF_fis97}) throughout most of the LWS 
range.  Single dust temperature models that fit our LWS spectrum with optical 
depth $\tau$(100\,$\mu$m) $\leq$ 1, would imply dust emission regions 3-4 times 
larger in angular area than what is measured at 1.3-mm by \cite*{JF_down98}, and 
more detailed radiative transfer models assuming central, spherical geometry, 
\cite*{JF_drainlee84} dust, and power law density point toward high optical depth even in the far infrared (Fischer et al. 1997; 2000).  Indeed, non-local
radiative transfer models of Eduardo Gonzalez-Alphonso and Pepe Cernicharo using 
identical ensembles of clouds with radii of 10$^{18}$ - 2 $\times$ 10$^{19}$ cm and densities of 10$^5$ - 10$^7$ cm$^{-3}$ can best explain the H$_2$O absorption lines in Arp 220 for $\tau_{FIR}$ $\geq$ 1 (\cite{JF_fis00}).  These model fits indicate that the levels are populated by absorption of photons emitted by warm dust.  This  suggests that the excited molecules may be located in PDRs associated with H~II regions or an AGN, where the infrared radiation field would be particularly intense. 

\cite*{JF_genz98} find that for Arp~220 a screen model with $A_v$ = 45 best fits the SWS mid-infrared recombination line fluxes, but point out that a mixed gas and dust model with $A_v$ = 1000 would provide moderate agreement in the near- and mid-IR.  If indeed $\tau$(150\,$\mu$m) = 1 and $A_v$ $\sim$ 1000 for 
Draine \& Lee dust (\cite{JF_fis97}), a purely mixed extinction model 
predicts an extinction correction of (1 - e$^{-\tau}$)/$\tau$ = 0.33 for the 
[\ion{O}{III}]88 \,$\mu$m line.  Even with $\tau$(150\,$\mu$m) = 10, the 88 \,$\mu$m line should have been detected at the 3$\sigma$ level for the pure mixed case.  While in the galaxies M~82 and Arp~299 the [\ion{O}{III}] 88\,$\mu$m line is about 1\% of the LWS far-infrared luminosity, it is $<$ 0.1\% (3$\sigma$) in Arp 220.  Thus, the extinction estimated from the mid-infrared lines is not enough to explain the lack of [\ion{O}{III}] 88\,$\mu$m line emission in Arp 220 if its intrinsic spectrum looks like that of M~82.  

Can low photoelectric heating efficiency in high $\langle$$G_o$$\rangle$/$n$ PDRs, as proposed by Malhotra et al. (1997; 2000), explain the low [\ion{C}{II}]  emission in ULIGs and other [\ion{C}{II}] deficient galaxies?  The [\ion{O}{I}] 63\,$\mu$m/[\ion{C}{II}] 158\,$\mu$m line ratio found in the two ULIGs where [\ion{O}{I}] 63\,$\mu$m is detected is consistent with higher $\langle$$G_o$$\rangle$/$n$ (\cite{JF_luhm00}). On the other hand, the upper limit on the [\ion{O}{I}] 145\,$\mu$m/[\ion{C}{II}] 158\,$\mu$m line ratio found in Arp 220 (Fischer et al. 1997) is inconsistent with high $\langle$$G_o$$\rangle$/$n$, unless there is  foreground absorption the [\ion{O}{I}]145 $\mu$m line, or if almost all of the [\ion{C}{II}] line emission in Arp~220 originates in the diffuse ionized medium, as suggested may be the case for many galaxies by \cite*{JF_malh00b}.  Comparing their observations of the ground state fine-structure line of atomic carbon to the [\ion{C}{II}] 158\,$\mu$m line, CO rotational lines, and the FIR flux, Gerin \& Phillips (1998; 2000) find evidence for high $\langle$$G_o$$\rangle$ and high $n$ in the ULIGs Arp~220 and Mkn~231, but suggest that dust absorption and/or [\ion{C}{II}] line opacity must also play a role.

\begin{figure}[tr]
  \begin{center}
    \epsfig{file=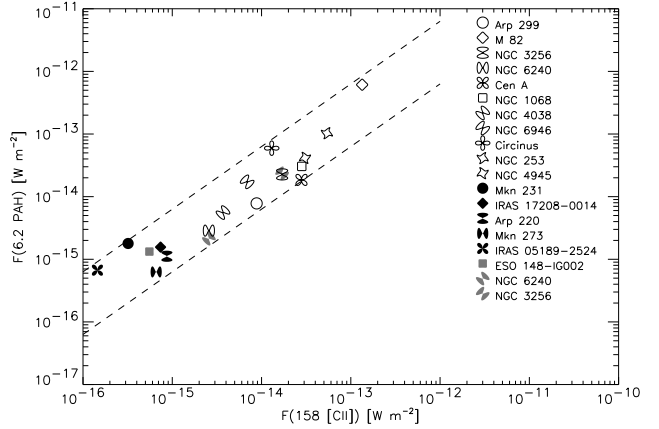, angle=90, width=8.5cm}
  \end{center}
\caption{ The 6.2 \,$\mu$m PAH feature versus versus [C II] 158\,$\mu$m line flux for ULIGs and other IR-bright galaxies.  The 6.2\,$\mu$m PAH emission was measured from CAM CVF spectra (open symbols), integrated over the full LWS beam for galaxies with extended emission, and from PHOT-S spectra (filled symbols) for compact galaxies (Dudley et~al. 2000). Black, filled symbols are ULIGs, while unfilled or grey filled symbols are lower luminosity galaxies.  Dashed lines show a range of a factor of 10 in the flux ratio.  From Luhman et~al. (2000). \label{JFfig_ciipah}}
\end{figure}

\cite*{JF_luhm00} have found that the average value and spread of the [\ion{C}{II}] 158\,$\mu$m line intensity to 6.2 \,$\mu$m PAH feature intensity ratio is similar in ULIGs and other IR-bright galaxies (Figure~\ref{JFfig_ciipah}).  The 6.2 \,$\mu$m feature was chosen because it is an isolated feature, well separated from other narrow PAH features and the 9.8 \,$\mu$m silicate absorption feature.  The 6.2\,$\mu$m PAH feature flux was measured from CAM CVF spectra, integrated over the full LWS beam for galaxies with extended emission, and from PHOT-S spectra for compact galaxies.  High $\langle$$G_o$$\rangle$/$n$ conditions causing low values of the [\ion{C}{II}]/FIR ratio, would not produce correspondingly low 6.2 \,$\mu$m feature flux, unless a significant fraction of the grains responsible for the 6.2 \,$\mu$m feature are destroyed by these conditions.  However, laboratory studies indicate that PAH sizes greater than 50 C-atoms probably dominate the emission in this band (\cite{JF_hudg99}) while theoretical studies indicate that it is difficult to destroy PAHs with greater than 50 C-atoms in the regimes of $\langle$$G_o$$\rangle$/$n$ thought to exist in these galaxies (Allain et~al. 1996).  The fact that similar flux ratios are found in both IR-bright galaxies and ULIGs therefore suggests that properties such as the average differential extinction and the average photoelectric heating efficiency are similar in these galaxies.  Alternatively, effects such as higher average differential extinction and lower photoelectric heating in ULIGs relative to other IR-bright galaxies may cancel each other to produce the observed linear correlation.

\cite*{JF_sco91} noted the low ratio of the infrared luminosity $L_{FIR}$ to the Lyman continuum photon rate $Q$ derived for Arp 220 from 2.7 mm observations.  In fact, based on the Lyman continuum rate measured at 3.3 mm for M~82 (\cite{JF_carlkron91}), and assuming $T_e$ = 10,000 K for both galaxies, $L/Q$ is a factor of 3.6 times higher in M~82. \cite*{JF_down98} suggest that this is due to increased dust absorption of Lyman continuum photons.  Comparison with CLOUDY (\cite{JF_ferl93}) models indicates that this can explain the low intensity of the FIR fine-structure lines, but would not produce lower excitation if \cite*{JF_drainlee84} dust is used.  Further modeling is needed to determine whether increased dust opacity within ionized regions, due perhaps to increased metallicity or evolutionary effects, can quantitatively explain the low [\ion{C}{II}]/FIR ratios, warmer dust temperatures, and OH and H$_2$O far-infrared absorption from excited levels. 

\section{SUMMARY}

Far-infrared spectroscopy obtained with the ISO LWS has shown that there is 
strong variation (more than 2 orders of magnitude) in the [\ion{C}{II}]/FIR ratios in galaxies extending from blue compact dwarfs, to normal and starburst galaxies, down to elliptical and ultraluminous galaxies.  The variation in the relative [\ion{C}{II}] line strength has been attributed to low metallicity 
(blue compact dwarfs), high $\langle$$G_o$$\rangle$/$n$ for normal galaxies and ULIGs, soft radiation fields in ellipticals, and extinction or enhanced abundance of dust in ionized regions in ULIGs. 

Full ISO/LWS far-infrared (43 - 197 $\mu$m) spectra of six nearby IR-bright galaxies reveal a dramatic progression of decreasing fine-structure line emission from ionized species to molecular absorption line dominated.  The archetypical ULIG, Arp~220 lies at the absorption line dominated end of this sequence.  For Arp~220, radiative transfer models indicate that it is optically thick in the FIR and that the water molecules observed in absorption are radiatively excited.  If extinction plays a role in the sequence it appears from this analysis that the affected regions are heavily obscured even in the far-infrared, while the \emph{detected} line emission is not more obscured in ULIGs than in starbursts.  Linear correlation between the PAH 6.2 $\mu$m feature strength and the [\ion{C}{II}] 158 $\mu$m line strength in starbursts and ULIGs suggests a similar effect for these emitting species, and that the detected PAH emission is not more obscured in ULIGs than in lower luminosity starbursts. 

Ultimately many of these effects are related to geometry and the physical 
location of dust, for which more detailed models and SIRTF spectroscopy will be 
invaluable.

\begin{acknowledgements}

I would like to thank my collaborators on the LWS infrared-bright galaxies science team, M. Luhman, S. Satyapal, C. Dudley, E. Gonzalez, P. Cernicharo, M. Wolfire, P. Clegg, J. Colbert, P. Cox, M. A. Greenhouse, V. Harvey, C. M. Bradford, S. Lord, M. Malkan, G. Melnick, G. Stacey, H. A. Smith, L. Spinoglio, G. Stacey, and S. Unger, for their contributions to the work reviewed here, some still in press or in preparation, and to S. Malhotra and S. Madden for access to preprints before publication.  I would also like to acknowledge the skill and dedication of the LWS instrument and data analysis teams at Vilspa, RAL, and IPAC.  This work was supported by the Office of Naval Research and the NASA ISO grant program.  

\end{acknowledgements}

\end{document}